\documentclass[runningheads]{llncs}
\usepackage{amsmath}
\usepackage{booktabs}
\usepackage[normalem]{ulem}
\usepackage{pgfplots}
\pgfplotsset{compat=1.11}
\usepackage{caption}
\usepackage[multiple]{footmisc}
\usepackage{subfig}
\usepackage{graphicx}
\usepackage{amsmath}
\usepackage{multirow}
\usepackage{amssymb}
\usepackage{xcolor}
\usepackage{comment}
\newlength{\bibitemsep}
\newlength{\bibparskip}
\let\oldthebibliography\thebibliography
\renewcommand\thebibliography[1]{%
  \oldthebibliography{#1}%
  \setlength{\parskip}{\bibitemsep}%
  \setlength{\itemsep}{\bibparskip}%
}

\begin{document}
\title{
Privacy-Aware Identity Cloning Detection based on Deep Forest\thanks{Please cite the paper as the following: \textbf{Alharbi, A., Dong, H., Yi, X, Abeysekara, P.: Privacy-Aware Identity Cloning Detection based on Deep Forest. The 19th International Conference on Service Oriented Computing (ICSOC 2021)}} 
%Privacy-Aware Social-Sensor Cloud Service Provider Identity Cloning Detection
}
\titlerunning{
Privacy-Aware Identity Cloning Detection based on Deep Forest
% Privacy-Aware SocSen Service Provider Identity Cloning Detection
}
\author{}
\institute{}
% If the paper title is too long for the running head, you can set
% an abbreviated paper title here
%
\author{Ahmed Alharbi\inst{} \and
Hai Dong\inst{} \and
Xun Yi\inst{} \and 
Prabath Abeysekara\inst{}
}
%
%\orcidID{2222--3333-4444-5555}
\authorrunning{A. Alharbi et al.}
% First names are abbreviated in the running head.
% If there are more than two authors, 'et al.' is used.
%
\institute{School of Computing Technologies, Centre for Cyber Security Research and Innovation, \\
RMIT University, Melbourne, Australia \\
Email: s3633361@student.rmit.edu.au,
hai.dong@rmit.edu.au,
xun.yi@rmit.edu.au, \\
s3693452@student.rmit.edu.au %\email{\{ahmed.alharbi,hai.dong,xun.yi,prabath.abeysekara\}@rmit.edu.au}
%\and
% \and
% \email
}
\maketitle   
%%\vspace{-1.3cm}% typeset the header of the contribution
\begin{abstract}
% Social sensing is a model that enables %\sout{multi-social}
% multiple social sensors such as humans, mobile phones, smart glasses to %\sout{crowd}
% source data. This sensed data can take various forms and be hosted %\sout{on}
% in social-sensor clouds (i.e. social networks) as social-sensor (SocSen) services. These services can be identified by their providers' social network accounts. Many attackers can infiltrate social-sensor clouds by cloning the user profiles of existing SocSen service providers to deceive social-sensor clouds. The %\sout{existing detection techniques}
% existing techniques to detect such infiltration mostly rely on privacy-sensitive user profile data, which are %\sout{inapplicable for}
% unavailable to be readily by the third-party applications. 
We propose a novel %\sout{SocSen service providers identity cloning detection method}
method to detect identity cloning of social-sensor cloud service providers to prevent the detrimental outcomes caused by identity deception. This approach leverages non-privacy-sensitive user profile data gathered from social networks and a powerful deep learning model to perform cloned identity detection. %\textcolor{red}{
We evaluated the proposed method against the state-of-the-art identity cloning detection techniques and the other popular identity deception detection models atop a real-world dataset. The results show that our method significantly outperforms these techniques/models in terms of Precision and F1-score.%}
\end{abstract}

%%\vspace{-1cm}
\section{Introduction}

\textit{Social sensing} is a model that enables %multi-\textit{social sensors} 
multiple \textit{social-sensors}, such as humans, smart phones and smart glasses, to gather data \cite{rosi2011social}. This sensed data, often referred to as \textit{social-sensor data}, can take various forms and be hosted on \textit{social-sensor clouds} (i.e. social networks, e.g. Twitter and Facebook) \cite{aamir2017social,aamir2017social1}. %\sout{For example, Facebook status messages and Twitter posts contain social sensor data}
Examples for such social-sensor data include Facebook status messages and Twitter posts. %\sout{These days}\sout{Nowadays} 
Social-sensor clouds are an important open medium that %\sout{expresses}
allows social-sensors/social media users to express their views on issues and events \cite{aamir2017social}. Critical information can be posted, especially descriptions and pictures of accidents or public activities \cite{rosi2011social}.
Social-sensor clouds currently play an important role in special events (e.g. sports, crimes, etc.). Thousands or even millions of posts can be published by social-sensors (in text and/or in images), over social-sensor clouds. This large amount of information can be summarised as \textit{social-sensor cloud services} (SocSen services) \cite{aamir2017social,aamir2017social1}. The special events can be represented  %defined 
from various points of view, such as where, when and what, by using the \textit{functional} and \textit{non-functional} properties of SocSen services \cite{9001217}.

%\sout{Social-sensor clouds} 
The proliferation of social-sensor clouds has drawn plenty of attackers in the recent past. These attackers often seek %\sout{seeking} 
to exploit the identities of \textit{SocSen service providers} (i.e. social media information providers) and deceive users in a variety of ways. One such method of exploiting SocSen service providers’ identities is identity cloning %\sout{which is}
in which an attacker registers a fake profile using the SocSen service provider’s identity information. %\sout{Identity}
Instances of identity cloning %\sout{is consists of}
can be divided into two types: single-site and cross-site identity cloning \cite{bilge2009all}. The former applies to cases in which an intruder establishes a cloned persona of a SocSen service provider in the same social-sensor cloud %\sout{. The} 
whereas the latter refers to cases %\sout{in which}
where an intruder steals a SocSen service provider's identity from another cloud. Here, we primarily focus on single-site identity cloning detection. 
% \begin{comment}
Many identity cloning related crimes have occurred in the past few years. %Examples %\sout{of related crimes such as}
%\textbf{include} tarnishing the reputation of the provider or %\sout{the friends of the provider}
%\textbf{their friends} to steal confidential information \cite{bilge2009all}. 
For instance, it has been reported that the Facebook Chief Executive Officer -  Mark Zuckerberg's Facebook account %\sout{, Chief Executive Officer of Facebook,} 
has been cloned and used in financial fraud\footnote{https://www.nytimes.com/2018/04/25/technology/fake-mark-zuckerberg-facebook.html}. %\sout{Another}
In another well-known %\sout{case is}
incident, the cloned Twitter account of Russian President Vladimir Putin%\sout{, which}
has attracted  %\sout{has} 
over 1 million followers\footnote{https://www.abc.net.au/news/2018-11-29/twitter-suspends-account-impersonating-vladimir-putin/10569064}.
The majority of social-sensor clouds does not support automatic detection of identity cloning. For example, Twitter and Instagram, %\sout{for example}
at present, investigate identity cloning reports after obtaining a valid identity cloning report from end-users. Automated methods to detect identity cloning are currently unavailable on these platforms\footnote{https://help.twitter.com/en/rules-and-policies/twitter-impersonation-policy}\footnote{https://help.instagram.com/446663175382270}. Meanwhile, %\sout{The} 
the majority of existing research on cloned identity detection uses \textit{both} privacy-sensitive and non-privacy-sensitive user profile attributes. Due to privacy limitations, third-party applications cannot access privacy-sensitive user profile attributes such as the user's full name, date of birth, or personal images available in social-sensor clouds through Application Programming Interfaces (APIs) or other means. %\sout{Consequently,}
As a result, most current techniques \cite{devmane2014detection,kamhoua2017preventing,kontaxis2011detecting} for detecting cloned identities are potentially less applicable to third-party applications. Suppose an intruder uses a cloned account to log into the web or application of a third-party. %\sout{This}
Then, by using current techniques, this third-party will have difficulty determining with certainty whether or not the account is cloned. %\sout{using current techniques}
In contrast, non-privacy-sensitive user profile attributes, such as the user's screen name, profile definition, and so on, are often readily available to third-party applications and can be directly accessed from %\sout{social networking platforms' APIs} 
the APIs exposed by social networking platforms. Hence, there is an %\sout{urgent}
apparent need and potential for exploring approaches to detect identity cloning by utilizing only non-privacy-sensitive user profile attributes. %In addition, most existing identity cloning approaches employ \textit{simple feature similarity} or \textit{machine learning models}. 
%\textbf{In addition, most existing identity cloning approaches only identify cloned profile by searching for profiles with simple feature (e.g. name, date of birth, etc) and they only return a ranked list of accounts that possibly have similar features to the victim account without identifying the correct cloned profile. Other approaches extract simple features for a given account pair. Then they fed the extracted features into a simple machine learning-based model. These models ignore the case where there are several cloned accounts of the same victim. Therefore, we need better alternatives that can detect all possible cloned profile.}

Moreover, the majority of current techniques detect cloned accounts using simple feature similarity \cite{devmane2014detection,goga2015doppelganger,kamhoua2017preventing,kontaxis2011detecting}. Simple feature similarity is typically calculated using human-defined metrics such as TF-IDF-based cosine similarity or Jaro–Winkler distance \cite{devmane2014detection,goga2015doppelganger,jin2011towards}. \textit{These metrics are incapable of encapsulating the semantics of a wide variety of literal strings} 
%\sout{. These metrics}
 and focus only on character%\sout{s} 
 distance or word frequency. %\sout{The above metrics, for example,}
For example, the above metrics cannot quantify the semantics of the terms \textit{king} and \textit{man }%\sout{, which define}
as a gender-based relationship. %\sout{Furthermore,}
In such a setting, deep learning (DL), as an emerging technology, has demonstrated its overwhelming performance on executing big data processing and analytics tasks in diverse fields \cite{najafabadi2015deep}. Nevertheless, to the best of our knowledge, using deep learning techniques to detect %\sout{the} 
identity cloning remains to be explored. Therefore, we intend to fill this gap by investigating how DL could potentially be applied in the domain of identity cloning.
%\sout{there has been no application of deep learning in identity cloning detection so far.}  
%\textbf{Explain the motivation (advantages compared to other DL/ML models) of deep forest here}
%Deep learning is approximately similar to deep neural networks (DNNs). Deep learning achieves great success in various applications and most of its applications are involving visual and speech information in recent years \cite{zhou2017deep}. 

Though some DL models are powerful, they are not directly applicable/suitable for our application on detecting identity cloning. The reasons are two-folds. %\sout{they have obvious disadvantages.}
Firstly, most of the DL models require a large amount of training data. This is required to effectively learn a representation of a phenomenon associated with the underlying training dataset. Secondly, %\sout{the hyperparameters of many DL models, such as deep neural networks (DNNs),}
most DL models such as Deep Neural Networks (DNNs) have many hyperparameters that need to be fine-tuned %\sout{, which can be a tedious and time consuming task \cite{zhou2017deep}. The}
 as the learning performance of most DNNs depends on how well their hyperparameters are configured. This can often be a tedious and time consuming task \cite{zhou2017deep}. In contrast, deep forest (DF) has a comparatively smaller number of hyperparameters that need tuning. 
Furthermore, %the ability of DF to automatically adjust the depth required seems particularly an enticing feature %\sout{self-adapting depth,} 
%by which the model complexity can be automatically determined.
the ability of DF to adaptively adjust the number of cascade levels required seems particularly an enticing feature by which the model complexity can be automatically determined. This enables DF to perform better on smaller data \cite{zhou2017deep}. %\sout{DF has much fewer hyperparameters compared to DNNs.} 
DF %\sout{also achieves}
has also been shown to achieve competitive performance compared to DNNs in %\sout{various types}
a variety of tasks \cite{zhou2017deep}.

To address the %\sout{above limitations}
limitations in existing identity cloning approaches outlined above, we formalize the identity cloning detection task as a classification problem using multiple representations of two accounts (account pair) as input. We propose a novel approach for %social media user
SocSen service provider identity cloning detection based on non-privacy-sensitive user information and a DF framework. %It consists of four main components: 1) a graph constructor (GC), 2) account pair feature representation, 3) multi-view account representation model and 4) \textbf{a} prediction model \textbf{based on DF framework}. 
Our main contributions can be summarized as follows:
\begin{itemize}
\item[--] We propose a novel SocSen service provider identity cloning detection method for third-party applications by utilizing only non-privacy-sensitive user profile attributes accessible through social-sensor cloud APIs.
%We designed a novel social media user identity cloning detection approach for third party applications, which depends only on non-privacy-sensitive user profile data accessed via social media APIs. 
%\item[--] We devised a user profile builder that constructs a unique user profile based on a set of features derived from non-privacy-sensitive user profile data. 
\item[--] We design multi-faceted representations to capture non-privacy-sensitive account features for effective identity cloning detection.
\item[--] We employ an effective DL model for cloned identify prediction. This is the first exploration on the application of DL in this %\sout{area}
problem setting. 
\item[--] We conducted extensive experiments using a real-world dataset. The experimental results show that our method produces higher precision and F1-score compared to the state-of-the-art identity cloning detection techniques, the other machine learning-based techniques and the variants of our proposed method.
%We collected a real-world dataset and conducted a set of experiments on it. NPS-AntiClone outperforms the state-of-the-art identity cloning detection approaches and other machine learning based approaches on Precision and F1-Score using non-privacy-sensitive user profile data. %\textbf{NPS-AntiClone detects identity cloning with a 96.0\% Precision, 76.03\% Recall and 84.86\% F1-Score.}
\end{itemize}
The rest of the paper is structured as follows. Section \ref{rw} reviews state-of-the-art identity cloning detection techniques. Section \ref{rw1} presents the details of our proposed solution. Section \ref{rw2} describes the evaluation %\sout{process} 
of our proposed solution and discusses the results obtained. Section \ref{rw3} concludes the paper.
%%\vspace{-0.5cm}
\section{Related work}\label{rw}
A significant number of techniques had been proposed in the current literature to detect spammers or fake accounts on social networks \cite{alharbi2021social}. The most commonly used techniques employ behavioural profiles %\sout{features}
of users that includes features such as writing style, accounts followed, etc, to classify users as trustworthy or untrustworthy \cite{masood2019spammer,zheng2015detecting}. However, the  aforementioned %\sout{Behavioural profiles} 
features %\sout{such as writing style, following accounts, etc}
used in behavioural profiles cannot detect cloned accounts accurately because an attacker attempting to clone the identities can imitate %\sout{the profile features of the victims}
those features. %\sout{Therefore, we need to}
This demands finding features that are able to accurately characterize account pairs to detect %\sout{the} 
cloned %\sout{identity}
identities. Meanwhile, %\sout{Other}
other works make use of %\sout{the} 
trust relationships that exist %\sout{s} 
between users in social networks to detect identity cloning. The main assumption of these methods is that a fake/spammer account cannot build an arbitrary number of trusted connections with legitimate accounts in social networks \cite{al2017sybil,masood2019spammer}. This assumption might not hold true in the context of identity cloning since attackers can attempt to imitate the profiles of legitimate. %\sout{accounts profile}
As a result, cloned accounts can build trust connections with legitimate accounts easier than other types of fake identities.

%\sout{In the current literature, a}
A few approaches have been proposed in the context of social media to %\sout{for detecting identity cloning on social media have been proposed}
detect identity cloning \cite{alharbi1234,alharbi2021social}. Kontaxis et al.  \cite{kontaxis2011detecting}  proposed a technique %\sout{for users to employ for determining}
that can be used by social medial users to determine if they have been a victim of identity cloning. Devmane and Rana \cite{devmane2014detection} devised a method %\sout{for}
to %\sout{detecting}
detect identity cloning in both single-site and cross-site contexts. To detect cloned profiles, %\sout{it}
the aforementioned approach searches for similar user-profiles and then computes a similarity index. Jin et al. \cite{jin2011towards}, in the meantime, analysed and characterised %\sout{the} 
identity cloning attacks' behaviours. They introduced two schemes for detecting suspicious profiles based on profile similarity. Kamhoua et al. \cite{kamhoua2017preventing} overcame identity cloning attacks by comparing user profiles across different social networks. They determined profile similarity using a hybrid string-matching similarity algorithm. Goga et al. \cite{goga2015doppelganger} proposed a method for detecting impersonation attacks. The proposed method determines whether two accounts are being used by the same person or an imposter. It first compares the behaviour and reputations of impersonation accounts %\sout{. It}
 and then detects impersonation attacks using a binary classifier trained using a Support Vector Machine (SVM). %\sout{binary classifier}.

%A few approaches had been proposed in the current literature to detect identity cloning on social media. Kontaxis et al. \cite{kontaxis2011detecting} proposed a methodology that can be employed by users to see if they have fallen victim to identity cloning. Devmane and Rana \cite{devmane2014detection} designed a methodology to detect identity cloning attacks in both single and cross-site contexts. It searches for similar user profile and then calculates a similarity index to detect the cloned profiles. Jin et al. \cite{jin2011towards} analysed and characterized the behaviours of the identity cloning attacks. They proposed two profile similarity schemes to detect suspicious profiles. Kamhoua et al. \cite{kamhoua2017preventing} overcame identity cloning attacks by matching user profiles across multiple social media. They used a hybrid string-matching similarity algorithm to find profile similarity. Goga et al. \cite{goga2015doppelganger} proposed an approach that can detect impersonation attacks.  Their approach detects whether a pair of profiles are controlled by the same person or imposter. It compares the impersonation account activity and reputations. It leverages a Support Vector Machine (SVM) binary  classifier  to detect impersonation attacks.

The majority of existing research \cite{devmane2014detection,kamhoua2017preventing,kontaxis2011detecting} detects identity cloning using both privacy-sensitive and non-privacy-sensitive user profile attributes. Many third-party applications and websites authenticate users from their social networking profiles. They are unable to access privacy-sensitive user profile attributes through social network APIs. Therefore, prior approaches may not be applicable to these third parties. In addition, the majority of the existing approaches \cite{devmane2014detection,goga2015doppelganger,jin2011towards,kamhoua2017preventing,kontaxis2011detecting} are built on simple feature similarity models or classic machine learning models. DL technologies have shown their superior performance on processing and analyzing big data in many application domains \cite{najafabadi2015deep}. To the best of our knowledge, there has been no attempt of applying DL technologies %\sout{to}
for identity cloning detection.
%%\vspace{-0.5cm}
\section{Methodology}
%%\vspace{-0.2cm}
\label{rw1}
In this section, we present a detailed overview of the proposed approach and its key components. 
%%\vspace{-0.5cm}
\subsection{Overview}
%%\vspace{-0.2cm}
%NPS-AntiClone aims to accurately detect identity cloning using non-privacy-sensitive user profile data. %It predominantly consists of three major components, namely, 1) a graph constructor that generate an undirected graph $UG$ to identify the pairs of similar social media user accounts, 2) an NPS-Profile Builder that constructs an NPS-Profile for a user account pair identified by the undirected graph from its non-privacy-sensitive user profile data, and 3) a DF model to predict if a user account pair contains a cloned user account.
%Fig. \ref{over} shows the workflow of NPS-AntiClone. 
%We first use a social media API to collect a set of user accounts and their non-privacy-sensitive user profile data. The graph constructor then builds an undirected graph from the user accounts to identify the pairs of similar social media user accounts. Next, the NPS-Profile Builder constructs an NPS-Profile for each paired account by defining and extracting two categories of non-privacy-sensitive user features from their respective original user profiles.
%\sout{the non-privacy-sensitive user profile data of each pair of accounts}. 
%These include similarity-based features, and differences-based features. These features allow analysing the similarities or differences between a given pair of accounts, as elaborated in Section 3.3. A DF model is designed to predict whether or not each account pair are an account and its replica using its NPS-Profile as the input. 

\begin{figure}[htbp]
  \centering
  \includegraphics[width=0.87\textwidth]{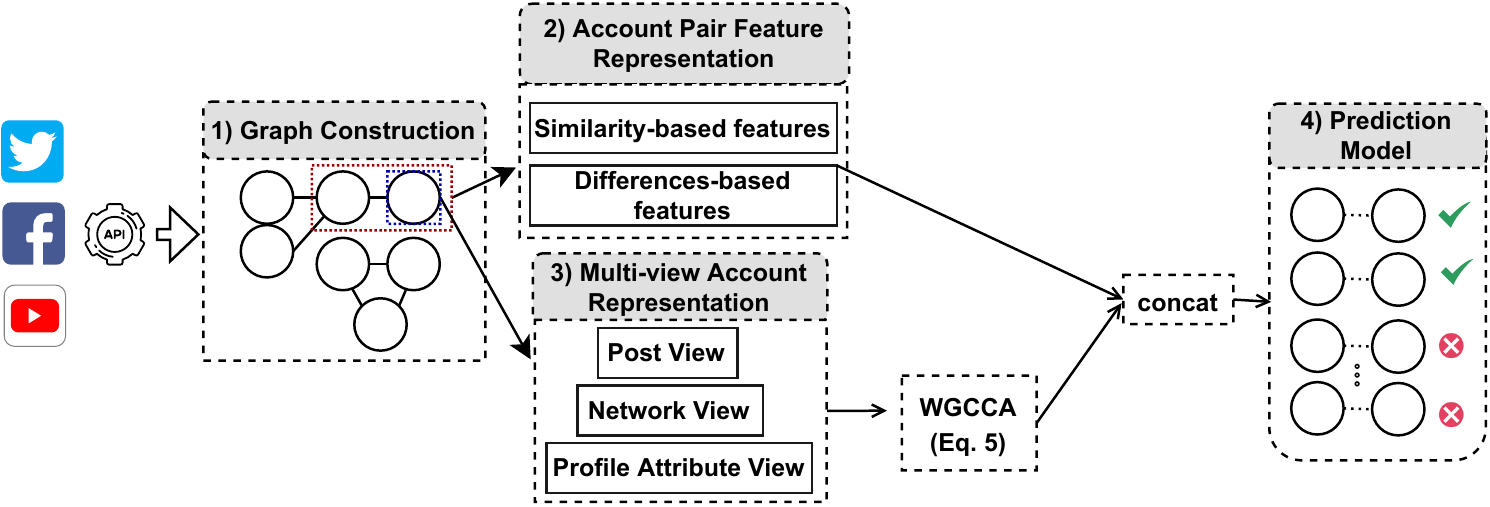}
  \caption{The workflow of our proposed methodology}
  \label{over}
\end{figure}

The proposed methodology %\sout{predominantly} 
is presented in Fig. \ref{over}, which consists of four major components, namely, 1) graph construction (GC) which aims to build an undirected graph from a given collection of social media accounts to identify the pairs of similar accounts; 2) %\sout{an}
an account pair feature representation, which extracts two categories of non-privacy-sensitive user features for each paired account; 3) a multi-view account representation, which constructs  a representation for each account  in  an account  pair from multiple non-privacy-sensitive perspectives; %given user (node) in the GC; 
4) a DF based prediction model, which predicts whether or not each account pair are an account and its replica using %\sout{the concatenation of}
a concatenated representation of the account pair feature %\sout{representation} 
and multi-view account representations. We discuss %\sout{the}
these four components in %\sout{more} 
detail in the following sections.

%%\vspace{-1cm}
%%\vspace{-0.5cm}
\subsection{Graph Construction}
%%\vspace{-0.2cm}
Given a set of social media accounts and their profile information, we aim to construct an undirected graph, where each pair of mutually connected nodes indicate the possibility that an account is the clone of the other.
%Then taking $UG$ as an input, \textbf{we aim to predict all the cloned accounts and their and its corresponding victim account}. 
%We aim to learn a predictive function $y = f(u_1,u_2)$, where $y$ is a binary value, with 1 indicating that a pair of user accounts identified by $u_1$ and $u_2$ contain a cloned account and its corresponding victim account, and 0 indicating otherwise. 
%We construct a graph to represent the similarity between accounts, where each node is an account. 
A cloned account is more likely to share the same screen name or username with the original account. Therefore, this graph connects nodes based on the screen name and username similarity. We connect two nodes with an edge only if the similarity score of the screen names or usernames of the two corresponding accounts is over a threshold $\delta$. This graph can locate almost all possible account pairs (i.e. a cloned account and its victim) in the dataset %\sout{,} 
while generating fewer false positives with an appropriate $\delta$ value. We %\sout{specify the}
elaborate as well as make recommendations on the process of determining an appropriate value of $\delta$ in the forthcoming experiments.
%\textbf{(needs to explain why choose 0.8)}, and we use the similarity score as the edge weight\textbf{(needs to explain how edge weight will be used later)}. An edge between two accounts does not mean that they have the potential possibility to form a pair of cloned and to be cloned account, but the true label is usually unknown.

%\subsection{NPS-Profile Builder}
%Once the undirected graph is constructed, NPS-Profile Builder will work on each pair of connected accounts.
%The NPS-Profile Builder focuses on extracting non-privacy-sensitive user features that can differentiate a cloned account from a non-cloned account, and create an intermediate user profile, termed an NPS-Profile. An NPS-Profile consists of two categories of non-privacy-sensitive user features, in the form of 
%1) similarity-based features and 2) differences-based features. Similarity-based features analyse the similarity between profile attributes of a pair of accounts (e.g., screen name, location, description). Differences-based features represent the difference between the general profile attributes of each account pair (e.g., number of tweets, number of followers).
%%\vspace{-0.5cm}
\subsection{Account Pair Feature Representation}
%%\vspace{-0.2cm}
Once the undirected graph is constructed, we then extract non-privacy-sensitive user features for each pair of connected accounts. The extracted features consist of two categories of non-privacy-sensitive user features %\sout{,} 
in the form of 1) similarity-based features and 2) difference%\sout{s}
-based features that can differentiate a cloned account from a legitimate account. We describe each of these feature(s) in detail in the following subsections.
%%\vspace{-0.5cm}
\subsubsection*{\textbf{Similarity-based features:}}These features are used to analyse the textual similarity between the non-privacy-sensitive attributes of the profiles that belong to a pair of accounts, such as username, screen name, description and location. Each feature is assigned a value from the range [0,1]. For example, when the username similarity feature is 1, this indicates the pair of accounts compared has 100\% textual similarity on the corresponding feature. In contrast, 0 indicates that the pair of accounts does not have any textual similarity on the given feature. We introduce the semantics of calculating the aforementioned textual similarity in more detail below.
%%\vspace{-0.3cm}
\paragraph*{Username, screen name and location similarity:} 
Previous studies have shown that the Jaro–Winkler string similarity (JS) performs best on the attributes' named values (e.g., username, screen name, or property name) \cite{christen2012data,cohen2003comparison}. Thus, we adopt JS, which is computed as the textual similarity between two strings as: 
    \begin{equation}\label{qwer}
   JS = \begin{cases}\frac{1}{3}. \frac{m}{\mid S1\mid}+ \frac{m}{\mid S2\mid} +  \frac{m-t}{\mid m \mid} & if :m>0\\0 & :otherwise \end{cases}
\end{equation}
where \(m\) is the number of matching characters, \(t\) is half the number of transpositions, and  \(\mid S1\mid\) and \(\mid S2\mid\) %\sout{is}
are the lengths of both strings. Matching characters are the same characters in the two strings with a maximum distance of \(w = \frac{max(\mid S1\mid, \mid S2\mid)}{2}\).
JS uses a prefix scale \(p\), which provides a more specific result when the two strings have a common prefix up to a specified maximum length \(l\).
        \begin{equation}\label{fl11}
        \begin{split}
   Jaro–Winkler = JS + p+l*(1- JS)
   \end{split}
\end{equation}
%For example, let us consider two strings `abs' and `sba', of which the number of matching characters is 3, but their order is not the same. Further,  the number of characters that are matching but not in order is 2, and accordingly, \(t\) is 1. Consequently, as per Eq. \ref{fl11}, the JS of the two aforementioned strings is 0.78.
%%\vspace{-0.3cm}
\paragraph*{Description similarity:} Users usually provide a short textual description of themselves %\sout{,} 
in their social media profiles, which commonly shows their associations to organizations, occupations and interests. Therefore, we calculate the description similarity of the account pair. We first pre-process the textual description by converting to lowercase, removing stop words and punctuation marks. We then use term frequency-inverse document frequency (TF-IDF) to convert the text description into vectors. %TF-IDF is a technique to compute a weight to each word based on how often the word appears in that corpus or document. It assigns a higher weight if the word appears more frequently, and a lower weight, if otherwise. %\cite{ramos2003using,salton1988term}.
%This technique is broadly adopted in text mining and information retrieval \cite{beel2016paper}. TF-IDF is defined against the following equation: 
 %   \begin{equation}\label{fl1331}
 %       \begin{split}
 %  TF-IDF (t,d) = TF(t,d) \times log (\frac{N}{df_t})
 %  \end{split}
%\end{equation}
%where \(t\) represents the term (word) in the account description, \(d\) is the account description (set of words), \(N\) is the total number of account descriptions, \(df_t\) is the count of occurrences of term \(t\) in the account description. %We derived a vector for each account pair description.%The set of the account pair descriptions is then viewed as a set of vectors. %in a vector space. Each term in the account pair description will have its own axis. 
We then used the cosine similarity to find the similarity between two account descriptions as:
\begin{equation}\label{fl13431}
        \begin{split}
   \cos(\theta) = \frac{\bf A\cdot\bf{B}}{||{\bf A}||\cdot||{\bf B}||}
   \end{split}
\end{equation}
where \(\bf A\) and \(\bf B\) are the TF-IDF scores of two accounts.
%and \(\bf B\) is the TF-IDF score of the description of the second account.
%%\vspace{-0.5cm}
\subsubsection*{\textbf{Differences-based features:}} These features are used to analyse the differences between the general profile attributes (e.g. the post count, friends count, etc.) that characterize individual accounts. We assume that the differences between the general profile attributes of a cloned account and its victim account will be higher than the other account pair. For example, a higher %\sout{tweets number}
degree of differences between the number of tweets can %\sout{be indicative of}
indicate an avatar %\sout{s} 
form of a pair of cloned and victim accounts.

Altogether, an account pair feature
% \sout{s} 
representation consists of 10 features across the two aforementioned categories, which are summarized in Table \ref{feature}.
%%\vspace{-0.5cm}
\begin{table}[t]
\caption{account pair feature representation and their descriptions}
\label{feature}
\resizebox{\textwidth}{!}{%
\begin{tabular}{p{0.32\textwidth}|l|p{0.3\textwidth}|p{0.75\textwidth}}
\toprule
\textbf{Feature category}& No. & \textbf{Features} & \textbf{Description} \\ \toprule
 \multirow{2}{*}{Similarity-based features}&1 & Username similarity  & Username similarity between the account pair. \\  
                  &2& Screen name similarity  & Screen name similarity between the account pair.  \\ 
                  &3& Location similarity &  Location similarity between the account pair.   \\ 
                  &4& Description similarity & Description similarity between the account pair.  \\ 
                  &5& Followers Ratio  & The ratio of the number of followers between the account pair. \\ \toprule

\multirow{2}{*}{Differences-based features}&6 &  Followers differences &  The number of followers difference between the account pair.  \\ 
                  &7& Friends differences  & The number of friends difference between the account pair.   \\
                  &8& Tweets differences   &  The number of tweets difference between the account pair.  \\ 
                  &9& Favorite differences  & The number of favorite difference between the account pair.   \\
                  &10& Account age differences  &  The account age difference between the account pair.  \\ 
                    \toprule
\end{tabular}%
}
\end{table}
%%\vspace{-0.9cm}
%%\vspace{-0.5cm}
\subsection{Multi-view Account Representation}
%%\vspace{-0.2cm}
Our objective is to construct a multi-view account representation %for a given user (node) in the GC 
for each account in the account pair by joining multiple views that correspond to the account's non-privacy-sensitive profile attributes. We utilise the post, network, and profile attribute views associated with the user account. These views can accurately reflect a user account, which attackers are highly likely to mimic. Then, a single embedding is learned from these views using weighted generalized canonical correlation analysis (wGCCA) \cite{hotelling1992relations}. Each view is discussed in detail in the following subsections.
%%\vspace{-0.5cm}
\subsubsection*{\textbf{Post View:}}
We obtain the pre-trained language representation for each account in the account pair %node (user) in the graph 
to generate the post view%first view
. We use the  Sentence-BERT (SBERT) \cite{reimers-2019-sentence-bert}  to obtain the user posts' vector-space representations. SBERT is an adjustment of the pre-trained bidirectional encoder representations from the transformers network (BERT) \cite{devlin2018bert}. 
%BERT is a widely used pre-training tool for language representations. It can be used to create high-quality language features or to fine-tune existing models for tasks such as question answering, classification, and so on \cite{devlin2018bert}. 
These pre-trained models are extremely efficient at extracting the text representation associated with any given task such as question answering, classification, %\sout{so on}
etc \cite{devlin2018bert}. %The reason for this is that BERT employs the Transformer architecture which is an attention mechanism that discovers contextual relationships between words in a text \cite{qiu2020pre}. 
SBERT generates semantically meaningful sentence representations using %\sout{siamese and triplet}
Siamese and Triplet network structures. %\sout{SBERT models, which is similar to the BERT models,}
Similar to BERT models, SBERT models are also based on transformer networks \cite{devlin2018bert}.  Additionally, SBERT performs a pooling operation on the output of BERT in order to obtain a sentence representation with a fixed length. Typically, the sentence representation is computed by calculating the mean of all output vectors. We collect $n$ posts that are publicly accessible for a given user account $u$,  denoted as $T = (t_1, ..., t_n)$. Each post $t_{i} $($i \in 1,..,n$)  is represented by the language representation that is pre-trained.
Each post $t_{i}$ is tokenized into a single word $w_{i}$ and then marked with special tokens called [CLS] and [SEP] to indicate the start and the end of a sentence, respectively. Then, a set of tokenized words is passed through BERT to embed fixed-sized sentences. Then, in the pooling layer, the $t$ representations are generated using mean aggregation. The mean aggregation is known to perform better than max aggregation or CLS aggregation \cite{reimers-2019-sentence-bert}. Each post's output is 768 dimensions, which is BERT's default setting. Finally, we aggregate all posts representation for the user account by computing the mean of all the posts' representation $T$.
%%\vspace{-0.5cm}
\subsubsection*{\textbf{Network View:}}
A network of accounts is a collection of users who interact in a variety of ways, such as friending, retweeting, and so on, within a social network, which can be represented as a graph. If one of the users in the social network interacts (i.e. follow, retweet, etc.) with another, an edge between them will appear in the graph. We consider two types of interaction networks: follower and friend networks. In the follower network, two users will be connected when a user follows a specific user (e.g. a friend or celebrity). In the friend network, two users will be connected when a user gets followed by another user. Inspired by the graph representation's success, we use the Node2Vec \cite{10.1145/2939672.2939754} to learn the network representation, or, in other words, the network view of an account. Node2vec is a widely used unsupervised graph representational learning technique. It utilises a biased random walk method to maximise the log-probability between a node's neighbours, or in other words, accounts with an edge between them. 
%%\vspace{-0.5cm}
\subsubsection*{\textbf{Profile  Attribute  View:}}
We obtain 12 non-privacy-sensitive user attributes to create an attribute vector to construct the profile attribute view of an account. These non-privacy-sensitive user attributes can be used to categorise an account's actions and credibility. For instance, the number of tweets may indicate a user's activity, while the number of followers may indicate a user's credibility. Table \ref{attributes} shows the 12 non-privacy-sensitive user attributes.  %We consider the following non-privacy-sensitive user attributes:
\begin{table}[t]
\caption{Features that profile attribute view contains and their descriptions }
\label{attributes}
\resizebox{\textwidth}{!}{%
\begin{tabular}{l|p{0.3\textwidth}|p{\textwidth}}
\toprule
 \textbf{No.} & \textbf{Features} & \textbf{Description}  \\ \toprule
1 & Friend (following) count & The number of accounts that the user follows.  \\ 
2 & Follower count & The number of followers that the account has.  \\ 
3 & Favorite count & The number of tweets liked by the account.  \\ 
4 & Tweet count & The number of tweets (including retweets) posted by the account.  \\ 
5 & List count & The number of public lists of which the account is a member.  \\ 
6 &  Account age & The account's lifetime to date, measured in months from the date of registration.  \\ 
7 & Profile background & A binary value indicating whether or not the account has changed the background or theme of their profile.  \\ 
8 & Profile image & A binary value indicating whether the account has not uploaded a profile image and instead uses the default image.  \\ 
9 &  Has profile description & A binary value indicating whether or not the account has added a description to their profile.  \\ 
10 &  Profile URL & A binary value indicating whether or not the account has added a URL to their profile.  \\ 
11 &  Screen name length & The length of the screen name of the account.  \\ 
12 &  Description length & The length of the description of the account.  \\ \toprule
\end{tabular}
}
\end{table}
%%\vspace{-0.5cm}
\subsubsection{Embedding Learning Model}
The proposed views in the previous section can include some helpful information that can be used to detect cloned accounts. Using each view separately can %\sout{result in a lack of}
cause a loss of valuable knowledge in comparison to using them in tandem. A simple and naive technique is to concatenate all the proposed views together. However, this concatenation might cause %\sout{an} 
over-fitting on %\sout{a} 
small training %\sout{sample}
datasets because of the resulting larger account representation or possibility of the resulting model %\sout{it may} 
ignoring the meaningful knowledge contained in the proposed views, as each view has %\sout{a} 
unique statistical %\sout{property}
properties. 
% Therefore, we employ canonical correlation analysis (CCA) which is a technique to learn single embedding from two views. %\sout{It}
% CCA extracts maximal information from the two views that are to be combined and generates a single embedding. CCA has been successfully applied to a variety of multi-view data learning problems. Since we generate more than two views in our proposed approach, we use Generalized CCA (GCCA), which allows multi-view based learning.
Therefore, we employ generalized canonical correlation analysis (GCCA) which is a technique to learn single embedding from multiple views.
% Multi-view based learning intends to learn a single function to represent multiple views and to optimize all the functions jointly to enhance their generalization performance \cite{zhao2017multi}. 
GCCA %\sout{consists of}
has many %\sout{variations}
variants, for example, \cite{carroll1968generalization,robinson1973generalized,tenenhaus2011regularized}. %[28], [29], and [30]. 
In our proposed approach, %\sout{We}
we use Carroll \cite{carroll1968generalization}’s GCCA %\sout{because}
as it is based on a computationally simple and efficient eigenequation. The GCCA objective can be formulated as follows:
\begin{equation}
\label{gcca}
    arg \min_{G_i,U_i} \sum_i \parallel G-X_iU_i \parallel_F^2 \qquad s.t. G' G = I 
\end{equation}
where \(X_i \in \mathbb{R}^{\ n\times d_i}\) corresponds to the data matrix %for 
of the $i^{th}$ view, $G \in \mathbb{R}^{\ n\times k}$ contains all learned account embedding 
%\textbf{%representation embedding} 
and $U_i \in \mathbb{R}^{\ d_i\times k}$ maps from the latent space to the observed view $i$. However, each view might have more or less knowledge for detecting identity cloning. As a result, we employ weighted GCCA (wGCCA). wGCCA adds weight %\textbf{s}
$w_i$ for each view $i$ %\sout{to}
in Equation \ref{gcca} as follows:
\begin{equation}
\label{wgcca}
    arg \min_{G_i,U_i} w_i\sum_i \parallel G-X_iU_i \parallel_F^2 \; s.t. G' G = I, w_i\geq0 
\end{equation}
where $w_i$ %represents the view's importance of the $i^{th}$ view. 
represent\textbf{s} the weight of a view and this weight shows the view's importance. The columns of $G$ are the eigenvectors of $\sum_i w_i X_i({X_i}'X_i)^{-1}{X_i}'$ and the solution for $U_i=({X_i}'X_i)^{-1}{X_i}'G$.
%%\vspace{-0.5cm}
\subsection{%The Deep Forest Model
Prediction Model}
%%\vspace{-0.2cm}
%We represented each account pair with their corresponding NPS-Profile. An account pair \(A_i\) can be represented by a feature vector \(F\), where \(F = (f_1, ..., f_n)\) (\(n = 10\)). 
%A pair of accounts \(A_i\) corresponding to a given NPS-Profile can be represented as a feature vector \(F\) ($\in \mathbb{R}^{10}$) where each feature $F_i$ in $F$ is indicative of an augmented feature derived based on the similarity or difference of a non-privacy-sensitive user profile feature.
%The DF model uses this representation to learn whether or not the account pair contains a cloned account and its corresponding victim account.
The final accounts pair representation \(A_i\) is the concatenation of the account pair feature representation and the multi-view account representation.
\begin{equation}
\label{concat}
    y = classifer(concat(F, wgcca) 
\end{equation}
where \(F\) is represented as a feature vector \(F\) ($\in \mathbb{R}^{10}$) where each feature $F_i$ in $F$ is indicative of an augmented feature derived based on the similarity or difference of a non-privacy-sensitive user profile feature and \(wgcca\) is the embeddings learned from the multi-view account representation. 

We employ DF to learn whether or not the account pair contains a cloned account and its corresponding victim account. The DF is a decision tree ensemble framework that can perform well even with relatively fewer data, and more importantly, has much fewer hyperparameters. The DF employs a cascade structure where each level of the cascade takes in a feature vector concatenated by its previous level and outputs its generating result to the next level of the cascade \cite{zhou2017deep}.

\begin{figure}[t]
  \centering
  \includegraphics[width=0.87\textwidth]{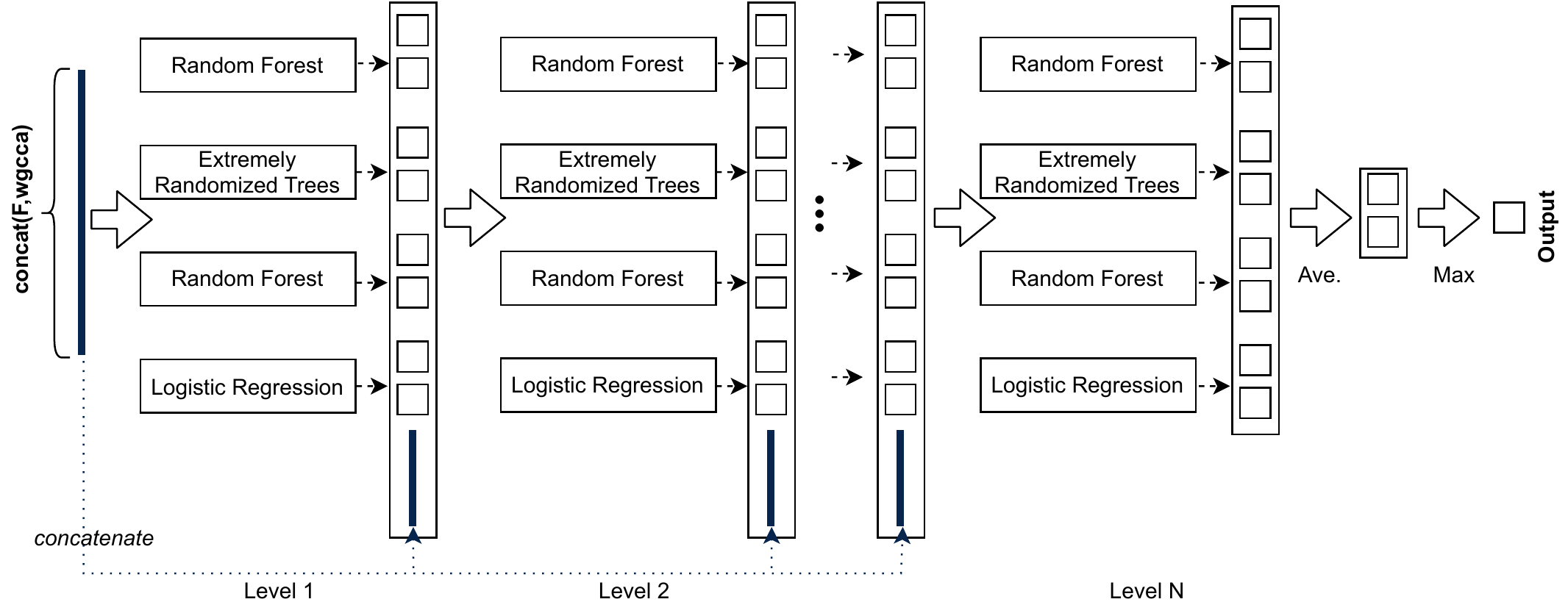}
  \caption{The architecture of the proposed deep forest model}
  \label{deep_forest}
\end{figure}

Diversity is highly recommended for ensemble construction \cite{zhou2012ensemble}. Therefore, the proposed DF model uses two types of random forests (RF), extremely randomized trees (ERT) and logistic regression (LR). Fig. \ref{deep_forest} shows the overall architecture of the proposed DF, which is of self-adapting depth with multiple levels. The RF model is an ensemble classifier which utilizes multiple decision trees at training time 
% \sout{on different subsamples} 
and uses averaging to get better prediction performance \cite{breiman2001random}. The ERT is similar to the RF model; however, they differ in the way splits are computed. The RF splits on trees while the ERT splits randomly \cite{geurts2006extremely}. 
%%\vspace{-0.5cm}
%%\vspace{-0.5cm}
%\textbf{LR is a statistical model that in its basic form employs a logistic function to model a binary dependent variable. It is also estimating the logistic model parameters \cite{wright1995logistic}.}

In the first level of the DF architecture, each model will take the $concat(F, wgcca)$ feature vector as an input %\sout{,} 
and %\sout{it then produces}
produce a class vector as output. This class vector is then concatenated with the $concat(F, wgcca)$ feature vector to be fed into the next level of the cascade. Herein, we aim to predict a binary value $y$ indicating whether an account pair contains a cloned account and its victim. Therefore, each of the four models (2 RFs, ERT, LR) produces a binary %\sout{classes}
output. Thus, the input of the next level of the cascade is composed of $8$ ($= 2\times4$) augmented features. %\sout{The} 
$k$-fold cross-validation $(k=5)$ is applied to generate the class vector. 
% More specifically, each model will be respectively trained %\sout{for} 
%  on $k-1$ splits of data. 
 This will result in $k-1$ class vectors that are then averaged to produce the final class vector as augmented features for the next level of the cascade. After expanding one more level, the prediction performance of the whole cascade is estimated by the validation set. The cascade growth is automatically terminated if there is no significant increase in the performance of class prediction%\sout{increase}
. Therefore, the number of cascade levels %\sout{is}
is said to be self-determined \cite{zhou2017deep}.
%%\vspace{-0.5cm}
\section{Evaluation}
\label{rw2}
%%\vspace{-0.2cm}
A set of experiments was performed to evaluate and analyse the effectiveness of our proposed solution against existing state-of-the-art identity cloning detection approaches. In addition, we also evaluated several candidate machine learning models to assess their performance in this particular problem context and justify the use of DF as the cloned identity predictor. %of NPS-AntiClone.
\begin{comment}

\textcolor{red}{All the experiments were conducted on a computer with an Intel Core i5 1.80 GHz CPU and 16 GB RAM. We used the gcForest package\footnote{https://github.com/kingfengji/gcForest} to implement the proposed DF model. We used the SBERT package\footnote{https://github.com/UKPLab/sentence-transformers} to extract the pre-trained language representations. We employed the StellarGraph package\footnote{https://github.com/stellargraph/stellargraph} to extract the Node2vec representations. We adopted Tensorflow %\footnote{https://www.tensorflow.org/} 
(v1.14.0) and scikit-learn %\footnote{https://scikit-learn.org/stable/} 
(v0.22.1) to implement the other DL and machine learning models. All the related works were re-implemented using Python and applied to our dataset.}
\end{comment}
All machine learning models were run for 10 rounds with different random permutations of the data. The results were presented as an average computed across the all rounds of experiments, together with standard deviation.
%\vspace{-0.5cm}
\subsection{Dataset}
%\vspace{-0.2cm}
%To the best of our knowledge, there is no readily and publicly available datasets that could be used for evaluating identity cloning in the domain of social networks. Most existing works, albeit limited, evaluated their proposed approaches on simulated data. Therefore, to evaluate NPS-AntiClone, we developed a dataset via authroized non-privacy sensitive user profile attributes fetched from Twitter APIs\footnote{https://developer.twitter.com/en/docs}. We collected 4,030 public Twitter accounts (2,015 cloned accounts and their corresponding victim accounts) from \footnote{https://impersonation.mpi-sws.org/}. We also randomly collected 20,152 %non-cloned 
%public Twitter accounts to add noises to the dataset. In total, we have 35,122 public Twitter accounts. The resulting dataset was randomly split with a 80:20 training-to-split ratio in order to derive training and test datasets.
To the best of our knowledge, %\sout{There}
there are no readily available and publicly accessible datasets for evaluating identity cloning detection in the context of social networks%\sout{ to the best of our knowledge}
. Most existing works, %\sout{although}
albeit limited, have used simulated data to evaluate their proposed techniques. We, therefore, developed a dataset via authorised non-privacy sensitive user profile attributes fetched out of Twitter APIs\footnote{https://developer.twitter.com/en/docs} in order to evaluate our proposed method. We collected 4,030 public Twitter accounts (2,015 cloned accounts and their corresponding victim accounts) from\footnote{https://impersonation.mpi-sws.org/}. We also randomly collected 20,152 public Twitter accounts to add noises to the dataset. Finally, we have 35,122 public Twitter accounts in total. The resulting dataset was randomly split with an 80:20 training-to-split ratio in order to derive training and test datasets.
%\vspace{-0.5cm}
\subsection{Other Approaches Evaluated}
%\vspace{-0.2cm}
% Alharbi et al.  has recently made a comprehensive survey on all the existing techniques in the context of the social media identity cloning.}Therefore, 
We compared and evaluated our proposed approach against the following existing state-of-the-art identity cloning detection approaches~\cite{alharbi2021social} and variants of the proposed approach.

\textbf{Basic Profile Similarity (BPS) \cite{jin2011towards}:} This technique determines the degree to which a given user profile and its suspected cloned account share public attributes and common friends. %Our experiments use only the number of common friends in the compared accounts' friend lists, as recommended and excluded friend lists are privacy-sensitive user profile attributes. %In our dataset, each account has 13  non-privacy-sensitive user attributes. We also set \(\mu = 0.0154\), \(\varepsilon = 13\), and \(\lambda = 0.03\). If the profile similarity of a given pair of accounts is less than $\mu$, then they are considered as non-cloned. For a given pair of accounts, $\mu$ is calculated based on the smallest number of the attribute similarity $a$ of the account pair and the smallest number of common friends $b$ of the account pair. $\varepsilon$ is the number of public attributes (raw data) in our dataset.  $\lambda$ is calculated based on the smallest $a$ and largest $b$ number of common friends in the account pair, respectively (see Equation-\ref{eq1234}).
    %\begin{equation}\label{eq1234}
    %\mu,\lambda = \frac{(a)^2+(b)^2}{\sqrt{k^2+x^2}}
%\end{equation}
%where $k$ and $x$ are adjusted based on the contribution of similar public attributes and the number of common friends.  We assigned 0.5 for $k$ and $x$ to make them contribute equally.

\textbf{Devmane and Rana \cite{devmane2014detection}:} This technique compares names, education, profile photos, places lived,  birthdate, workplace, gender, photos added to the profile, and number of friends/connections. %We only compared the screen name, places lived (location) and the number of friends/connections that can be fetched using Twitter APIs. We then calculated a similarity index for a given pair of accounts. The original work does not show adequate details on the type of similarity index they used in their proposed approach.  Therefore,  we adopted the screen name and location similarity technique used in our work to calculate the similarity index in the compared user profiles.

\textbf{Goga et al. \cite{goga2015doppelganger}:} This method compares 
% the dates of account creation and the accounts' popularity and social influence to determine whether an account pair contains an impersonated account. It uses four distinct types of features, namely 
profile similarity, social neighbourhood overlap, time overlap accounts, and account differences. It then trains an SVM classifier using a linear kernel to determine whether or not a given account has been impersonated.
    %This technique compares the time at which the pair of accounts were created and the reputations of the accounts in terms of popularity and social influence, and detects whether an account pair has an impersonated account. To determine this, it uses four different types of features, namely,  profile similarity, social neighbourhood overlap, time overlap accounts and differences between accounts. It then trains an SVM classifier, with a linear kernel to classify if a given account is impersonated. %In our experiments, we used all the  recommended features in the original work to train the SVM classifier. 

\textbf{Kamhoua et al. \cite{kamhoua2017preventing}:} This method compares friend list similarity and calculates attribute similarity using a modified similarity metric  called Fuzzy-Sim. To calculate the attribute similarity, it considers the following attributes: name, education, city, age, workplace, gender, and friend list. We used the same threshold values recommended by the original work (0.565 and 0.575) for the Fuzzy-Sim.   %In our experiments, however,  we only compared the screen name, city (location), and the friend list provisioned by Twitter APIs. We then calculated the attribute  similarities and friend list similarity using FuzzySim  with the same threshold values recommended by the original work (0.565 and 0.575).

\textbf{Zheng et al. \cite{zheng2015detecting}:} This is a typical model for detecting spammers. It makes use of 18 features, some of which are profile-related, such as the number of followers, and others of which are content-related, such as the average number of hashtags. It then trains an SVM classifier using a Radial Basis Function (RBF) kernel to determine whether an account belongs to a spammer or not.% In our experiments, we used all the recommended features recommended in the original work to train the SVM classifier.

\textbf{GC ($\delta=0.8$):} This is a variant of our proposed solution that only feeds the results of the graph construction into  the DF model.

\textbf{Account:} This is also a variant of our proposed solution that only feeds the account pair feature representation into the DF model. %${Account}$. %Additionally, we attempt to use the similarity-based features and differences-based features separately and denote them ${SIM}$ and ${DIF}$ respectively.

\textbf{WGCCA:} This is also a variant of our proposed solution that only feeds the multi-view account representation into the DF model. %Additionally, we attempt to use the similarity-based features and differences-based features separately with DF$_{WGCCA}$ and denote them ${WGCCASIM}$ and ${WGCCADIF}$ respectively.

We further compared our proposed DF model against %the GC and 
the following machine learning and DL models in order to justify the use of the DF as the identity cloning predictor. These models are broadly applied in the context of social media identity deception detection \cite{alharbi2021social}. %in NPS-AntiClone. 
These models are, namely, Adaboost (ADA), Convolutional Neural Network (CNN), Deep Neural Network (DNN), K nearest neighbours (KNN), LR, Multi-layer Perceptron (MLP) and RF.  %Support vector machine (SVM), 
In addition, we compared the proposed DF model with the other types of DF models (RF-based DF (DF$_{RF}$), ERT-based DF (DF$_{ERT}$) and LR-based DF (DF$_{LR}$)) to further justify the performance of the model. %All the model hyperparameters have been properly tuned to achieve their optimal performance. Table \ref{parm} shows the hyper-parameter values used for machine learning and DL algorithms evaluated. We also find the optimal $\delta$ value of the graph constructor is 0.8 according to the experimental results. 
%%\vspace{-0.5cm}
\begin{table}[t]
\centering
\caption{Hyperparameter values used for the candidate machine learning and DL algorithms}
\label{parm}
\resizebox{\textwidth}{!}{%
\begin{tabular}{p{0.3\textwidth}p{0.9\textwidth}}
\toprule
\textbf{Model} & \textbf{Parameter} \\ \toprule
ADA & estimators = 50 \\
CNN & 10 layers, filters = 64, kernel size = 2, pool size = 2 \\
DNN & 5 layers (250, 200, 50, 1)\\
KNN & neighbors = 5 \\ 
MLP & solver = adam, activation = relu \\
RF & estimators = 50 \\
%SVM & kernel = linear \\ 
\toprule
\end{tabular}%
}
\end{table}
% %\vspace{-0.5cm}

%We also compared NPS-AntiClone with the variants of non-privacy-sensitive user features to evaluate the impact of each category of non-privacy user features.
%We used similarity-based features and differences-based features separately.
%\vspace{-0.5cm}
\subsection{Hyperparameter Tuning}
%\vspace{-0.2cm}
All the hyperparameters of the supervised machine learning models were properly tuned to obtain their optimal performance. Table \ref{parm} shows the hyperparameter values used for the machine learning configured and (or) tuned the parameters %\sout{following their}
as recommended in their respective original works. We also only used the non-privacy-sensitive user attributes provisioned by Twitter APIs. For our proposed solution, we find the optimal $\delta$ value of the GC is 0.8 according to the experimental results.
% (described in Section. 4.4). %We set the parameters of the state-of-the-art approaches following the original works. 
We used `paraphrase-distilroberta-base-v1'\footnote{https://huggingface.co/sentence-transformers/paraphrase-distilroberta-base-v1} as the pre-trained model for SBERT. This pre-trained model was trained on millions of paraphrase sentences. %\sout{The dimension size of the post representation using SBERT is $768$, which is a default setting of the model}
SBERT by default uses $768$ as the dimension of its post representation. The default dimension %\sout{size} 
of the Node2vec for both the follower and friend network is $128$. We also used the probability %\sout{for}
of moving away from source node $q = 2$, the probability of returning to source node $p = 0.5$, the number of random walks per root node $n= 10$ and the maximum length of a random walk %\sout{is}
as $15$. All the profile attribute views were normalized to %\sout{range 0 and 1}
$[0, 1]$. The weights $w$ of the wGCCA %\sout{is}
were set as $[0.25, 0.5, 0.5, 0.25]$ via experiments.
%\vspace{-0.5cm}
\subsection{Results and Discussion}
%\vspace{-0.2cm}
\begin{comment}
\textcolor{red}{We evaluated and compared the performance of all the aforementioned models based on the mainstream performance metrics adopted 
in the area of identity cloning detection: Precision, Recall and F1-score \cite{alharbi2021social}. Precision indicates the percentage of predicted account pairs that are correctly detected (including a cloned account and its victim). Recall indicates the proportion of true account pairs correctly detected. F1-score is the harmonic mean of precision and recall.}
\end{comment}
%\vspace{-0.4cm}
\paragraph*{\textbf{Overall performance:}} The performance comparison results are shown in Table \ref{zzz}. Our proposed solution %\sout{yields}
yielded the best %\sout{performing results}
amongst the models compared %\sout{on}
against Precision and F1-Score. BSP \cite{jin2011towards} achieved %\sout{s}
 the second %\sout{best-performing technique}
best performance %\sout{on}
against Precision and F1-Score. Meanwhile, Goga et al. \cite{goga2015doppelganger}'s proposed technique yielded %\sout{s}
 the best-performing result on Recall. BSB \cite{jin2011towards} only employs profile attribute similarities and shared friends between account pairs. Goga et al. \cite{goga2015doppelganger}'s proposed technique compares an account pair using only a traditional similarity technique and does not consider a filtering method for the account pair similar to our GC. The methods introduced by Kamhoua et al. \cite{kamhoua2017preventing} and Devmane and Rana \cite{devmane2014detection} use a simple method to calculate profile attribute similarity. These techniques do not take into account the impact of the account's posts and the account's network information representation. Zheng  et al. \cite{zheng2015detecting}’s proposed technique only takes into account a subset of features that compares spammer behaviour patterns. The obtained results indicate that our proposed solution is more suitable %adaptable 
to %\sout{the context that all the user profile data employed for identity cloning detection is the non-privacy-sensitive user profile data provisioned by social media APIs}
a setting where only the non-privacy-sensitive user profile attributes are used for identity cloning detection in social media.

%The results of our experiments show that our proposed approach clearly outperforms the existing state-of-the-art identity detection techniques on precision and F1-Score (see Table \ref{zzz}). Our proposed approach achieved a Precision of 96.02\%, Recall of 76.03\%, and F1-Score of 84.86\%, respectively, which were 20\%, 1\% and 8\% higher than that of the BSP \cite{jin2011towards}, the second best-performing approach.  The identity detection approaches proposed by Kamhoua et al. \cite{kamhoua2017preventing} and Devmane and Rana \cite{devmane2014detection} performed poorly because they use lower numbers of non-privacy-sensitive user features. BSP, on the other hand, depends on the number of public attributes (raw data) and common friends between the cloned and non-cloned accounts. It does not consider the effect of the engineered features such as username similarity and location similarity, etc. The obtained results indicate that NPS-AntiClone is more adaptable to the context that all the user profile data employed for identity cloning detection is the non-privacy-sensitive user profile data provisioned by social media APIs.

\begin{table*}[t]
\centering
\caption{
Comparison with the state of the art identity cloning approaches. Standard deviation ($\sigma$) is provided to the KPIs of our proposed solution and Goga et al.'s algorithm that were evaluated over 10 iterations.
}
\label{zzz}
\resizebox{\textwidth}{!}{%
\begin{tabular}{p{0.4\textwidth}p{0.25\textwidth}p{0.25\textwidth}p{0.25\textwidth}}
\toprule
\textbf{Model} &   \textbf{Precision ($\sigma$)} & \textbf{Recall ($\sigma$)} & \textbf{F1-Score ($\sigma$)}  \\ \toprule
BSP \cite{jin2011towards} & 76.8 & 75.1 & 76.9    \\
Devmane and Rana \cite{devmane2014detection}  & 66.3 & 68.1  & 67.2   \\
Goga et al. \cite{goga2015doppelganger} & 65.4 (1.1) & \textbf{85.9 (1.5)} & 74.3 (0.7)   \\ 
Kamhoua et al. \cite{kamhoua2017preventing}  & 68.2 & 70.1  & 69.1  \\
Zheng et al. \cite{zheng2015detecting} & 68.15  & 73.34   & 70.64  \\
Our proposed solution    &  \textbf{90.08 (3.42)} &	77.95 (1.69) &	\textbf{83.52 (0.94)}  
 \\ 
\toprule
\end{tabular}%
}
\end{table*}

The comparison of results among the proposed DF model %\sout{ and}
as well as %\sout{the} 
other machine learning and DL models revealed that our DF model significantly outperformed all the other candidate models (see Table \ref{zzz1}). We attribute this superior performance to two reasons. 
%We attribute this significant increase in performance of NPS-AntiClone to the following two key characteristics of DF. %namely, 1) the rich diversity characteristics it promotes via the use of multiple different types of base-learners (i.e. RF, ELT, LR) in ensembles and 2) its ability to adaptively determine the complexity of the predictive model in a data-dependent way. 
First, our DF model generates an ensemble of its base learners (i.e. RF, ERT, and LR) with a cascading structure where each cascade is an ensemble of the aforementioned base learners. Consequently, such a behaviour encourages diversity thereby improving its generalization performance eventually contributing to the significant increase in performance of our proposed solution. Further, the ability of DF to adaptively determine the best model complexity required for the problem context at hand also allows generating comparatively simpler models compared to the other approaches such as DNNs. This also helps improve the generalization performance of DF \cite{zhou2017deep}.
%\textbf{The reason behind it is that DF generates a deep forest ensemble, with a cascade structure. Each cascade is an ensemble of decision tree forests (an ensemble of ensembles). NPS-AntiClone has different types of models (i.e. 2 RFs, ERT, LR) and this encourages the diversity which is crucial for ensemble construction \cite{zhou2012ensemble}. This cascade structure enables the DF to do representation learning. Another reason is that it employs a DL which adaptively decides its model complexity by terminating training when adequate. This enables it to be applied to different scales of training data, not limited to large-scale ones. In contrast, most deep neural networks complexity is fixed and requires large scale training data \cite{zhou2017deep}.}%\sout{The reason behind it is that NPS-AntiClone employs a DL which adaptively decides its model complexity by terminating training when adequate \cite{zhou2017deep}. This enables it to be applied to different scales of training data, not limited to large-scale ones. In contrast, most deep neural networks complexity is fixed.}
% %\vspace{-0.5cm}
\begin{table}[t]
\centering
\caption{Comparison with the baseline machine and DL models evaluated as the predictor of our proposed solution. %NPS-AntiClone. 
Each KPI is presented as an average over 10 iterations together with standard deviation (\(\sigma\)).}
\label{zzz1}
\resizebox{\textwidth}{!}{%
\begin{tabular}{p{0.4\textwidth}p{0.25\textwidth}p{0.25\textwidth}p{0.25\textwidth}}
\toprule
\textbf{Model} &   \textbf{Precision (\(\sigma\))} & \textbf{Recall (\(\sigma\))} & \textbf{F1-Score (\(\sigma\))}  \\ \toprule
%$GC$ ($\delta=0.8$) & 83.02 & 76.36  & 79.55    \\
ADA   & 84.63 (1.49) & 78.15 (1.49)  & 81.26 (0) \\ 
CNN   & 83.54 (0.41) & 77.67 (1.87)  & 80.49 (1.20) \\ 
DNN  & 85.66 (2.38) & 76.68 (2.12)& 80.88 (0.42)  \\ 
KNN  & 78.30 (1.49) & 78.52 (0) & 78.41 (1.49)   \\ 
LR & 84.64 (0) & 78.15 (1.49) & 81.27 (1.49)   \\
MLP  & 86.55 (3.81) & 72.55 (3.69) & 78.83 (2.37)   \\ 
RF & 83.84 (0.27) & 78.56 (0.12) & 81.12 (0.12) \\  \hline
%SVM & 81.37 (2.01)  & 76.13 (0.92) & 78.66 (1.40) \\ \hline
DF\(_{RF}\) & 84.16 (0.49) & \textbf{78.82 (0.25)}	 & 81.41 (0.34)	\\
DF\(_{ERT}\) &  83.98 (0.16) & 78.44 (0.25)	 & 81.11 (0.06)	\\
DF\(_{LR}\) & 80.42 (0.433) & 77.64 (1.32)	 & 79.01 (0.77)	\\
DF  & \textbf{90.08 (3.42)} &	77.95 (1.69) &	\textbf{83.52 (0.94)} \\ \hline
%DF_{SIM} & 86.31 (1.58) & 74.06 (0.73)	 & 79.70 (0.44)	\\
%DF_{DIF} & 90.08 (1.65) & 74.47 (0.67) & 81.53 (0.62)	\\
\toprule
\end{tabular}%
}
\end{table}
\section{Conclusion and Future Work}
\label{rw3}
%\vspace{-0.2cm}
We propose a novel SocSen service provider identity cloning detection approach based on non-privacy-sensitive user attributes. In the proposed approach, %\sout{An}
an undirected graph is first constructed to identify the pairs of similar SocSen service providers’ identities. We then extract %\sout{ed} 
an account pair feature representation. %\sout{We}
Afterwards, we also extract %\sout{ed for each account}
a multi-view account representation for each account. Then, these two representations are concatenated and fed into a DF classifier to predict whether or not a given account pair contains a cloned account. Our proposed solution was evaluated on a real-world Twitter dataset against other state-of-the-art cloned identity detection techniques and machine learning models. The results show that the proposed approach significantly outperformed the other models. 

In the future, we plan to develop more effective identity cloning detection techniques and conduct experiments on large-scale datasets.
% a mechanism for determining which account of an account pair is the cloned account, since %\sout{our proposed solution currently}
% the solution proposed in this work only detects if an account pair consists of a pair of cloned accounts and their victim.
%\vspace{-0.5cm}
%\frame[shrink=50] {\printbibliography} 
%\printbibliography
%\bibliographystyle{ieeetr}
%\bibliographystyle{splncs85}
\bibliographystyle{splncs04}
\bibliography{ref.bib}

%\bibliographystyle{splncs04}
%\bibliography{/ref.bib}

\end{document}